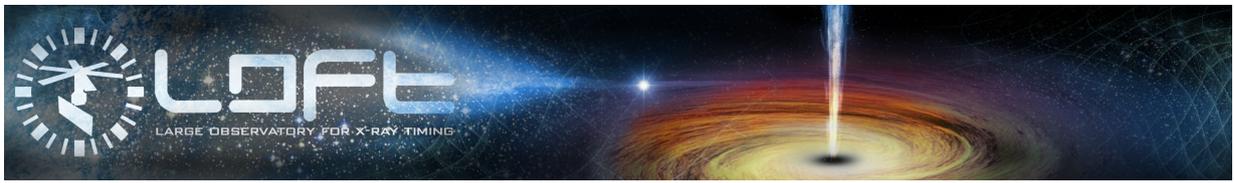

# Accretion, ejection and reprocessing in supermassive black holes

## White Paper in Support of the Mission Concept of the Large Observatory for X-ray Timing

### Authors

A. De Rosa[1], S. Bianchi[2], M. Giroletti[3], A. Marinucci[2], M. Paolillo[4,5,6], I. Papadakis[7,8], G. Risaliti[9], F. Tombesi[10,11], M. Cappi[12], B. Czerny[13,14], M. Dadina[12], B. De Marco[15], M. Dovciak[16], V. Karas[16], D. Kunneriath[16], F. La Franca[2], A. Manousakis[13], A. Markowitz[18], I. McHardy[17], M. Mehdipour[19], G. Miniutti[20], S. Paltani[21], M. Perez-Torres[22], B. Peterson[23], E. Piconcelli[24], P. O. Petrucci[25,26], G. Ponti[15], A. Rozanska[13], M. Salvati[9], M. Sobolewska[13], J. Svoboda[16], D. Trevese[27], P. Uttley[28], F. Vagnetti[29], S. Vaughan[30], C. Vignali[31], F. H. Vincent[13], A. Zdziarski[13]

[1] National Institute for Astrophysics/IAPS. Rome. Italy
[2] Università degli studi Roma Tre. Rome, Italy
[3] National Institute for Astrophysics/IRA. Bologna, Italy
[4] Università degli studi di Napoli Federico II. Naples, Italy
[5] ASI Science Data Center, Rome, Italy
[6] INFN - Sez.di Napoli. Naples, Italy
[7] University of Crete, GR-71003 Heraklion, Greece
[8] IESL, Foundation for Research and Technology, GR-71110 Heraklion, Greece
[9] National Institute for Astrophysics/Observatory of Arcetri. Florence, Italy
[10] University of Maryland. USA
[11] NASA - Goddard Space Flight Center, USA
[12] National Institute for Astrophysics/IASF-Bologna. Bologna, Italy
[13] N. Copernicus Astronomical Center PAS, Warsaw, Poland
[14] Center for Theoretical Physics, Warsaw, Poland
[15] Max-Plack Inst. for Extraterrestrial Physics, Germany
[16] Astronomical Institute, Academy of Sciences, Prague, Czech Republic
[17] University of Southampton. Southampton, United Kingdom
[18] University of California, San Diego, La Jolla, CA, USA
[19] SRON Netherlands Institute for Space Research, Utrecht, Netherlands
[20] Centro de Astrobiología (CSIC-INTA), Dep. de Astrofísica, Madrid, Spain
[21] University of Geneva. Geneva, Switzerland
[22] IAA-CSIC. Granada, Spain
[23] Ohio State University, Columbus, OH, USA
[24] National Institute for Astrophysics/Observatory of Rome. Rome, Italy
[25] Université Grenoble Alpes, IPAG, F-38000 Grenoble, France
[26] CNRS, IPAG, F-38000 Grenoble, France.
[27] Università di Roma La Sapienza. Rome, Italy
[28] University of Amsterdam, Amsterdam, Netherlands
[29] Universitá di Roma Tor Vergata. Rome, Italy
[30] University of Leicester, Leicester, United Kingdom
[31] Università di Bologna. Bologna, Italy

## Preamble

The Large Observatory for X-ray Timing, *LOFT*, is designed to perform fast X-ray timing and spectroscopy with uniquely large throughput (Feroci et al. , 2014). *LOFT* focuses on two fundamental questions of ESA's Cosmic Vision Theme "Matter under extreme conditions": what is the equation of state of ultra-dense matter in neutron stars? Does matter orbiting close to the event horizon follow the predictions of general relativity? These goals are elaborated in the mission Yellow Book (http://sci.esa.int/loft/53447-loft-yellow-book/) describing the *LOFT* mission as proposed in M3, which closely resembles the *LOFT* mission now being proposed for M4.

The extensive assessment study of *LOFT* as ESA's M3 mission candidate demonstrates the high level of maturity and the technical feasibility of the mission, as well as the scientific importance of its unique core science goals. For this reason, the *LOFT* development has been continued, aiming at the new M4 launch opportunity, for which the M3 science goals have been confirmed. The unprecedentedly large effective area, large grasp, and spectroscopic capabilities of *LOFT*'s instruments make the mission capable of state-of-the-art science not only for its core science case, but also for many other open questions in astrophysics.

*LOFT*'s primary instrument is the Large Area Detector (LAD), a $8.5\,\mathrm{m}^2$ instrument operating in the 2–30 keV energy range, which will revolutionise studies of Galactic and extragalactic X-ray sources down to their fundamental time scales. The mission also features a Wide Field Monitor (WFM), which in the 2–50 keV range simultaneously observes more than a third of the sky at any time, detecting objects down to mCrab fluxes and providing data with excellent timing and spectral resolution. Additionally, the mission is equipped with an on-board alert system for the detection and rapid broadcasting to the ground of celestial bright and fast outbursts of X-rays (particularly, Gamma-ray Bursts).

This paper is one of twelve White Papers that illustrate the unique potential of *LOFT* as an X-ray observatory in a variety of astrophysical fields in addition to the core science.

# 1 Summary

*LOFT* will make a decisive step forward towards our understanding of the inner structure of Active Galactic Nuclei (AGN) thanks to the huge effective area and the CCD-class energy resolution of the LAD, coupled with the large sky coverage of the WFM. The good control of the background systematics ($\lesssim 0.15\%$ of the average background level) will allow the LAD to reach a sensitivity limit for S/N $\sim$ 3(250) of $0.03(3) \times 10^{-11}$ erg cm$^{-2}$ s$^{-1}$ (2–10 keV), and therefore observe about 100 AGN with a very good accuracy (S/N $\sim$ 100) and more than 300 AGN with a S/N level of a few tens. Moreover, thanks to its large sky coverage, the WFM will be capable of continuous X-ray monitoring of large numbers (>100) of AGN on time scales of a few days to weeks. The WFM will thus provide an essential synergy with those observatories that will, in parallel to LOFT, be opening up AGN time domain science on those timescales and at other wavebands, e.g., the Cherenkov Telescope Array (*CTA* - TeV energy domain), *Meerkat*/Square Kilometre Array (*SKA* - radio wavelength) and the Large Synoptic Survey Telescope (*LSST* - optical band). Here, we focus on three key aspects of AGN physics (related to accretion, ejection and reprocessing in Supermassive Black Holes), which will be specifically addressed by the unique capabilities of *LOFT*.

- *Investigating the central engine through its variability.* **LOFT** can revolutionize our view of the X-ray variability in AGN, which is thought to originate in the innermost regions of the accretion flow. The crucial advancement of the LAD instrument, with respect to the old and existing X-ray detectors, is the unprecedented low level of the Poisson noise level in its light curves. LAD observations will therefore offer an accurate determination of the characteristics of the Power Spectral Density (PSD) functions for the AGN population in the local Universe, thus allowing the study of their correlation with physical parameters like Black Hole (BH) mass, luminosity, and accretion rate. Moreover, its enormous collecting area will allow us to study in detail the dependence of the AGN X-ray variability on energy, the only possible way to test conclusively Comptonization flare models. This will provide us with direct insight into the nature of the primary emission region, i.e., the size, the density, and optical depth, the exact dissipation process (shocks or magnetic field reconnection), and also the location of the event from its own individual reflection component. WFM observations can also play a crucial role in our understanding of the X-ray variability in AGN. This instrument will survey large areas of the sky, allowing both individual (for the brightest) and ensemble studies of long-term variability for several hundreds of AGN. Thanks to the wide fraction of accessible sky available with the WFM and the high sensitivity for short exposures available with the LAD, *LOFT* will be an ideal partner for *SKA* in the physics of accretion in AGN.

- *Mapping the circumnuclear region around the central Supermassive BH.* The standard "torus" invoked by Unification Models, in the generic sense of an axisymmetric, rather than spherically symmetric, circumnuclear absorber is certainly present in AGN, but its physical and geometrical structure is far from being one and the same for all the objects. The most effective way to estimate the distance and properties of the absorber is by means of the analysis of the variability of its column density $N_H$. In particular, the LAD will observe rapid (from a few hours to a few days) $N_H$ variability in bright AGN with significantly higher precision with respect to current X-ray telescopes. Moreover, the broad band offered by *LOFT* will allow us to explore in detail those clouds with larger column densities, whose variability mostly affects the spectrum at high energies. These occultations may also produce variations of the profile of the relativistic iron K$\alpha$ line, allowing a precise determination of the BH spin, and an unambiguous probe of general relativistic effects in the strong-field regime, one of the primary scientific objectives of *LOFT*. Meanwhile, the synergy between the WFM standard operation, and a well-designed monitoring campaign with the LAD, will allow us to take advantage, for the first time, of X-ray reverberation mapping of the narrow core of the Fe K$\alpha$ emission line, and therefore estimate the geometry and distance of the material

- *The BH-galaxy interaction: ultra-fast ionized winds.* The high effective area and CCD-like energy resolution of *LOFT* LAD can also provide a very important contribution to the understanding of the co-evolution between the supermassive BH and its host galaxy. In particular, powerful ultra-fast outflows (UFOs) with velocities in excess of 0.1c, likely identified with accretion disc winds or the base of a possible weak/broad jet, could provide a significant contribution to AGN "feedback". The unique combination of very high effective area peaking at ∼10 keV, moderate energy resolution, and continuous coverage up to energies of ∼20–30 keV will allow a breakthrough in UFO studies compared to current and upcoming X-ray satellites. Indeed, *LOFT* LAD will be able to observe these transient features in the relevant timescale of variability (few ks) and, thanks to the broad energy band, will allow us to characterize fast outflows (0.6–0.7$c$) in local sources, which are possibly associated with the highest mechanical power, the ones with the highest impact on AGN feedback.

## 2 Introduction

The standard model of AGN predicts a supermassive black hole (BH), surrounded by a geometrically thin, optically thick accretion disc. The UV/optical seed photons from the disc are Compton scattered into the X-ray band through a hot corona above the accretion disc which produces the hard power-law spectrum extending to energies determined by the electron temperature in the hot corona (with the cut-off energy ranging around ∼100–200 keV, Perola et al. 2002, De Rosa et al. 2012, Molina et al. 2013, Brenneman et al. 2014, Marinucci et al. 2014, Ballantyne et al. 2014). This primary X-ray radiation in turn illuminates the inner disc and it is partly reflected towards the observer's line of sight (Haardt & Maraschi 1991, Haardt et al. 1994), producing an iron line that will be heavily distorted, with a pronounced red wing, a clear signature of strong field gravity effects on the emitted radiation. This "inner structure" is surrounded by rapidly rotating clouds of gas, the Broad Line Region (BLR). A toroidal absorber is located at a distance of ∼ a parsec from the BH and within the Narrow Line Region (NLR) clouds. These regions contribute with additional features to the primary continuum: a narrow Fe K line at 6.4 keV and a reflection hump above 10 keV, both due to the reprocessing of the continuum from the optically-thick medium (such as the torus and/or the accretion disc), and the absorption and variable features produced in the ionized and cold material around the inner region.

However, in the last decade X-ray observations of AGN have called for a revision of the simplest configuration of the Unified Model (e.g., Bianchi et al. 2012). The luminosity and variability of the central source play a role in the amount of absorbing matter and their properties (Della Ceca et al. 2008). In fact, photoionized warm absorbers (WA) are present in at least 50% of unobscured AGN (Crenshaw et al. 2003, Blustin et al. 2005), and about 35% of the bright AGN show high-density, highly-ionized fast outflowing winds ($v \sim 0.1$–$0.15c$; Tombesi et al. 2010, Gofford et al. 2013). This strongly suggests either the presence of more than one absorbing/emitting region, likely outflowing, or a single large-scale stratified outflow observed at different locations from the BH. In addition, the evidence of a correlation between radio and X-ray emission in AGN has suggested that the inflowing and outflowing components near the BH are coupled, although the detailed physical/radiative mechanisms at work are still a matter of study (Panessa et al. 2013, Orienti et al. 2014). It is then fundamental to be able to characterize the physical properties of the sub-pc cores together with the incidence of associated jet/outflow structures in AGN.

The huge effective area of the LAD, with a good control of the background systematics and CCD-class energy resolution, together with the large sky coverage of the WFM onboard *LOFT*, will make a decisive step forward in our understanding of the inner structure of the AGN using spectral/timing information.

The LAD onboard *LOFT* is a collimated and non-imaging instrument, hence its background cannot be simultaneously measured, and then subtracted during the source observation, as usually happens in imaging instruments. Instead, it has to be carefully modelled and calibrated "a priori" (Campana et al. 2013). Detailed simulations of the instrument show that the current *LOFT*/LAD design is expected to reach a background rate not higher than $6 \times 10^{-11}$ erg cm$^{-2}$ s$^{-1}$ in 2–10 keV, and about $2 \times 10^{-10}$ erg cm$^{-2}$ s$^{-1}$ in the 2–30 keV energy band. The largest fraction of the background intensity is due to the so-called leakage component, i.e., the Cosmic X-ray background photons that pass through collimator shields. Its intrinsic nature allows then for a very efficient modelling during the orbital period, with an estimated systematic residuals value of 0.15%, which is significantly better than the requirement value of 0.25%.

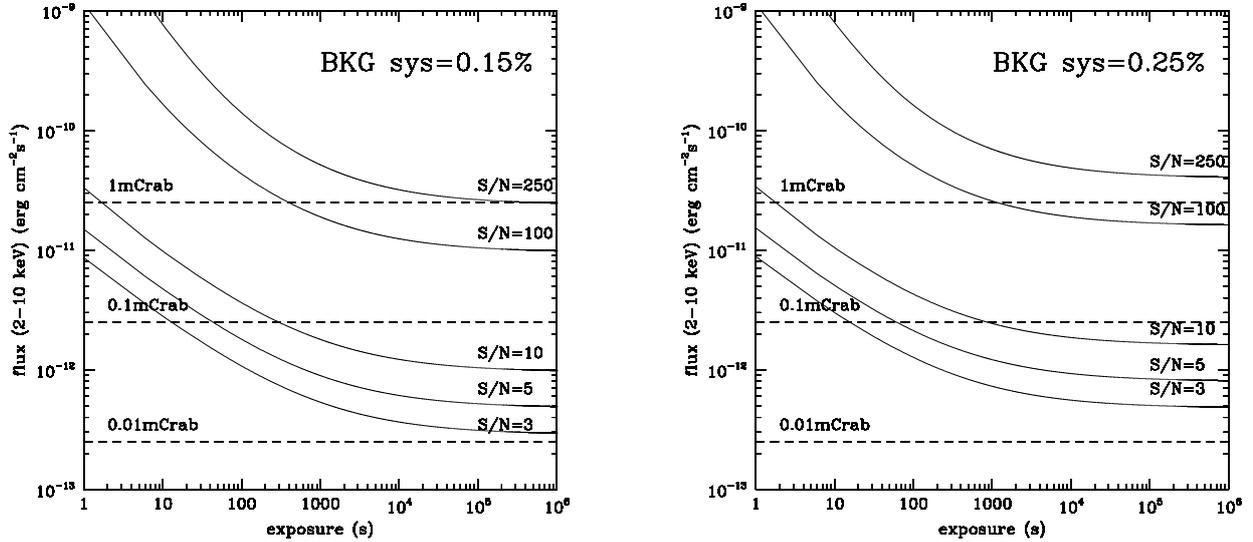

Figure 1: LAD sensitivity curves for S/N = 3, 5, 10, 100 and 250. On the left we assume the estimated value for the background systematics, 0.15%, while on the right the requirement value of 0.25%.

In Fig. 1 we show the sensitivity curves of the LAD for values of the S/N ranging from 3 to 250 in 2–10 keV and different levels of background systematics (0.15%, 0.25% of the mean background level), constant over the energy band. The effect due to the uncorrected background variations (background systematics) will add an extra statistical noise component, significantly increasing the limiting flux attainable for each given S/N. The sensitivity limit for S/N∼3 (250) ranges from 0.03 (3) to 0.05(4) $\times 10^{-11}$ erg cm$^{-2}$ s$^{-1}$ for 0.15% and 0.25% of the background systematics, respectively.

Thanks to its large sky coverage (field of view of 4.1 sr), the broad energy range (2–50 keV), good position accuracy (1′), and its sensitivity (3$\sigma$ in 1 day being about $\sim 10^{-10}$ erg cm$^{-2}$ s$^{-1}$ in the full band 2–50 keV) the WFM will play a crucial role in our understanding of the X-ray variability in AGN: it will be capable of providing X-ray monitoring and observation of large numbers (>100) of AGN on daily timescales. The WFM sky survey during the *LOFT* lifetime will open the investigation for X-ray variability of AGN on different time scales, providing a powerful tool to investigate the emission properties of supermassive BH in their innermost region. It is important to stress that there will be observatories which are opening up the AGN time domain on those timescales for the first time in other wavebands and for which simultaneous X-ray monitoring will be critical, e.g., *CTA* (TeV), *Meerkat/SKA* (radio) and *LSST* (optical).

In this paper we present the results obtained from detailed simulations to investigate the capabilities of both the LAD and WFM to advance our knowledge on AGN. These results clearly demonstrate the potential of *LOFT* to increase significantly our knowledge on the accretion, ejection and reprocessing processes in AGN

## 3 Investigating the central engine through its variability

### 3.1 X-ray variability in the frequency domain

X-ray variability in accreting compact objects is thought to originate in the innermost regions of the accretion flow and can provide valuable information regarding the nature and the geometry of the X-ray emitting source. Our knowledge of the X-ray variability in AGN has advanced substantially in the last 15 years, thanks mainly to *RXTE* and day-long *XMM-Newton* observations of a few, X-ray bright Seyferts. AGN show "red noise" power spectral density functions (PSD) that decreases steeply at high frequencies (short timescales) as a power law, PSD($\nu$) $\propto \nu^{-\alpha}$ (where $\nu$ is the temporal frequency), typically with $\alpha \sim 2$. Below some characteristic frequency, $\nu_b$, the PSDs flatten, and these bend frequencies scale approximately inversely with the BH mass from BH-XRBs to AGN (e.g., McHardy et al. 2006). There is an indication that, for a given BH mass, these bend frequencies increase with increasing accretion rate, although this is not fully established yet (e.g., Gonzalez-Martin & Vaughan 2012). This is a crucial issue: although all the relevant accretion disc timescales do depend on the mass of the central object, only a few of them may depend (indirectly) on the accretion rate as well. So far, our knowledge is based on the power-spectral analysis of just ~20 objects. In addition, in some cases the bend frequencies are not well determined yet (estimates are basically just lower limits) while the analysis has been done using data from different satellites, and applying different methods and models to the data. Due to its unprecedented effective area and the significant improvement in the M4 background systematics, LAD observations will allow for vast improvement in the determination of the X-ray PSDs of AGN. As a result, *LOFT* can potentially revolutionize our view of the X-ray variability in AGN.

Since *LOFT* will be a low Earth orbit satellite, observations will consist of data segments with a duration of ~ 3 ksec, interrupted by Earth occultation periods. Despite this uneven sampling pattern, accurate PSD estimates are still possible, if one uses the short period segments to estimate the PSDs to frequencies as low as $(3\,\mathrm{ks})^{-1}$, and then bin the lightcurves with a bin of size equal to the orbital period (e.g., Papadakis et al. 2002). As a result, a 300 ks observation of an AGN will typically result in a light curve which will consist of ~ $100 \times 3$ ks segments, and will have a total duration of ~500 ks. The left panel in Fig. 2 shows the PSDs of an AGN with a $10^5$, $10^6$, $10^7$ and $10^8\,M_\odot$ BH mass (solid, dashed, dot-dashed, and dotted lines, respectively), assuming a "bending-power law" model, with a high-frequency slope of $-2$, which "bends" to a slope of $-1$ below $\nu_b$. The bend frequency is assumed to scale with BH mass and the PSD amplitude is equal to the current estimate of the average PSD amplitude (Gonzalez-Martin & Vaughan 2012). The PSDs are shown over the frequency range that will be sampled by a typical LAD observation, i.e., $2 \times 10^{-6}$–$9.3 \times 10^{-5}$ Hz, using the orbital period binned light curve, and $3.3 \times 10^{-4}$ up to 0.1 Hz (assuming bins of size 5 s for the short segments).

The crucial advancement of the LAD instrument, compared to the previous and existing X-ray detectors, is the unprecedented low level of the Poisson level in its light curves. The expected Poisson level is indicated with the thick, horizontal red line in the same figure, and is estimated based on the fact that LAD will offer a $3\sigma$ detection in just 1 sec for a flux of $\sim 10^{-11}$ erg cm$^{-2}$ s$^{-1}$ source. The low Poisson power level will allow an extremely accurate determination of the *high* frequency PSD slope, which is crucial for estimating $\nu_b$, especially for objects with a mass lower than $\sim 10^7\,M_\odot$. Overall, the resulting broad-band LAD PSDs will offer a very accurate characterization of the PSD bend frequency, and PSD slopes both in the high and low frequencies (at least for objects with a BH mass as large as $\sim 5 \times 10^7\,M_\odot$). Since there are more than 400 AGN with a flux equal or larger than $10^{-11}$ erg cm$^{-2}$ s$^{-1}$ (Revnivtsev et al. 2004), LAD observations will offer an accurate determination of PSD characteristics for the AGN population in the local Universe, thus allowing the study of their *distribution* and their correlation with BH mass, luminosity, spectral shape etc.

LAD's enormous collecting area will also allow us to study in detail the dependence of the AGN X-ray variability on energy. LAD will provide a ~ $3.5\sigma$ detection of a ~ 0.04 mCrab source in the 2–10 keV band over a time period of just 100 s. This unprecedented capability will allow, for the first time, to characterize accurately the X-ray PSDs at these energies even with a relatively low exposure time (a 100 ks exposure will be enough to determine accurately the high frequency PSD) and study in detail the PSD bend frequency, amplitude and slope evolution with energy. Only such a study can test conclusively Comptonization flare models, as they predict different timing evolution in different energy bands (Malzac & Jourdain 2000), and the possibility of an outflowing corona in AGN (Malzac et al, 2001). In addition, accurate determination of the PSD at high energies can also provide well sampled Fourier-resolved spectra, which can determine accurately the variable spectral components in AGN, and test conclusively the presence of variable absorbers near the central source (Arévalo & Markowitz 2014).

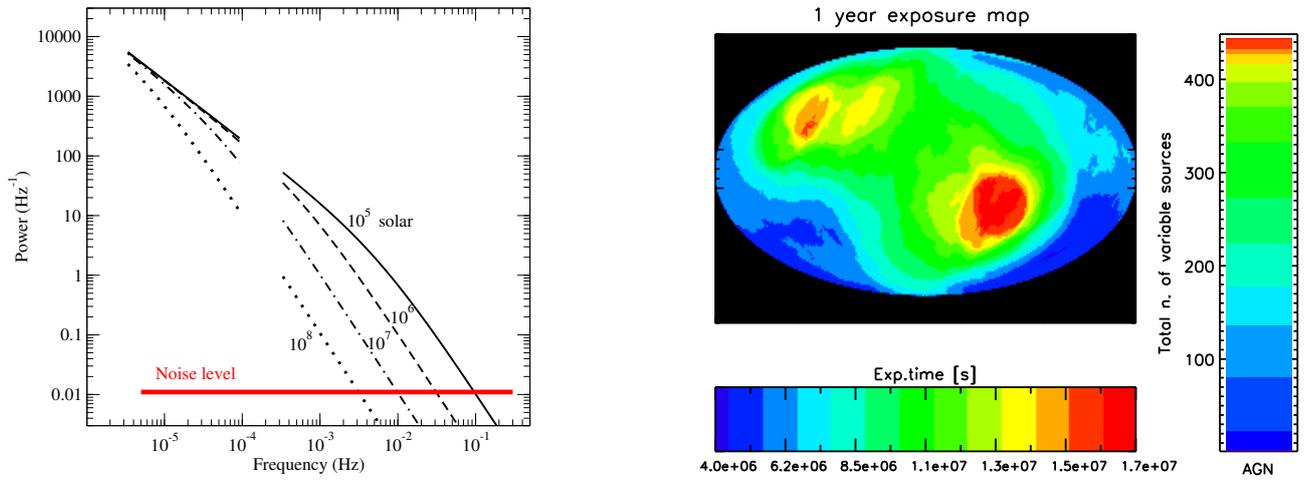

Figure 2: *Left:* PSDs of an AGN with a BH mass of $10^5$, $10^6$, $10^7$ and $10^8 M_\odot$ (solid, dashed, dot-dashed, and dotted lines, respectively), assuming a "bending-power law" model, with a high-frequency slope of $-2$, which "bends" to a slope of $-1$ below $\nu_b$. The PSDs are shown over the frequency range that will be sampled by a typical LAD observation. The expected Poisson level is indicated with the thick, horizontal red line, and is estimated based on the LAD sensitivity ($3\sigma$ detection in 1 sec for a $\sim 10^{-11}$ erg cm$^{-2}$ s$^{-1}$). *Right:* Expected 1 year WFM exposure time in seconds as a function of the sky position (Galactic coordinates in Aitoff projection). Only locations covered with at least 10 cm$^2$ of total net detector area are considered. The bottom colorbar shows the net exposure times, while the vertical colorbar shows the total number of AGN for which the WFM will return light curves with at least 10 points and where variability will be detected with S/N > 3.

## 3.2 The multi-wavelength context in 2025: X-ray, radio and optical variability

### 3.2.1 WFM and black hole mass estimates

Wide Field Monitor observations, with a typical $5\sigma$ sensitivity in 1 day of about $2 \times 10^{-10}$ erg cm$^{-2}$ s$^{-1}$, can also play a crucial role in our understanding of the X-ray variability in AGN. This instrument will survey the sky both in "triggered" and in "normal" operating mode, the latter being a low timing and spectral resolution mode active when no triggering source is present. In this mode, the WFM will survey large areas of the sky allowing to study long-term X-ray variability of bright/nearby AGN. To demonstrate the unique capabilities of the WFM we used the 1-year exposure sky map shown in the right panel of Fig. 2 as a good approximation of the adopted pointing strategy. Using the AGN number counts observed by previous missions, we computed the number of sources which will be accessible to *LOFT* in each region of the sky, based on the M4 sensitivity estimates (assuming an average AGN spectrum and Galactic absorption). Using the current "typical" AGN PSD, we estimated the number of sources which are bright enough to build a lightcurve with a minimum of 10 time bins (in the 1-year period) and for which variability will be detected with S/N > 3. To be conservative we assumed the instrument to be operating in "normal" low resolution mode. The color map on the right panel of Fig. 2 shows the total number of AGN which fulfill these criteria in each region of the sky. Overall we estimate that, even in its first year of operations, WFM will return light curves with at least 10 points in them, for several hundreds of *variable* AGN, allowing both individual (for the brightest) and ensemble studies of AGN variability on long (months) timescales.

Allevato et al. (2013) demonstrated that it is possible to retrieve the intrinsic variability (within a factor of 2) even when using sparse lightcurves with low S/N, such as the ones expected here for most of the faint AGN population in the WFM sample. Depending on the exact monitoring pattern, much lower uncertainties are expected for the brightest sources and on averaging samples of >20 sources. We will therefore be able to determine accurately the *intrinsic* variability amplitude (on long time scales) for hundreds of AGN, and study the variability-luminosity relation with an unprecedented accuracy. This relation will then be used to compare with similar relations for AGN at higher redshifts, and also to acquire BH mass estimates for the objects in the sample. Although the uncertainty of the BH mass estimates may be large, on an individual basis, the estimates for the whole sample will be useful to construct the BH mass (and accretion rate) distribution for the local AGN population. The serendipitous sky monitoring by the WFM will allow us to significantly measure the variability (excess variance) of hundreds of local ($z < 0.3$) AGN. As already discussed, in the last years several studies have found a significant anti-correlation between the BH mass and the X-ray variability (O'Neill et al. 2005; McHardy et al. 2006; Gierlinski et al. 2008; Zhou et al. 2010; Ponti et al. 2012; Kelly et al. 2013). The variability measures by the WFM, together with the PSD measures by the LAD (see Fig. 2), will allow us to obtain a first direct estimate of the local BH mass function of the AGN. This measurement will be very important to constrain the AGN evolutionary scenarios. In particular, while at high masses the local BH mass function is expected to be mainly dependent on the merging history of galaxies, at low mass it is instead able to provide important upper limits on the primordial BH mass function (Volonteri & Bellovary 2012).

In fact, one of the most important issues on the AGN evolution is the measurement of the BH mass function and how it has evolved with time. However up to now the BH mass function is derived only indirectly by using the BH mass–bulge scale relations which are calibrated on only a few tens of local AGN and should be used with caution for low BH masses ($< 10^6$ $M_\odot$; e.g., Ferrarese & Merritt 2000; Magorrian et al. 1998; Marconi & Hunt 2003). In addition, the well known method of BH mass estimation through optical reverberation mapping (Peterson et al. 2004) seems, somehow, selective. In fact, this method imposes the presence of BLR visible along the line of sight, then it is not applicable to type 2 AGN. We also note that optical monitoring could be not always suitable for mass estimates, being the origin of optical variability very complex and the scaling of variability with BH mass still unclear, although it has been proposed it may scale with the monochromatic luminosity (see Czerny et al. 2011). However, the BH mass determination from the X-ray excess variance is simpler (one observation of several ksec in X-rays instead of 100 ksec in optical well distributed in time for several months or so) and less biased (in optical there is the inclination angle of the system surely involved, while X-rays are likely isotropic enough).

### 3.2.2 *LOFT* and the Square Kilometre Array (*SKA*)

There is a long debate about the existence of a correlation between radio and X-ray emission in radio quiet AGN, with important implications on the physics of accretion onto supermassive BH. Various authors proposed a correlation between radio and X-ray luminosity, with different slopes and involving different parameters, including the BH mass (Merloni et al. 2013, Wang et al. 2006, Panessa et al. 2013); other authors, (e.g., Burlon et al. 2013 using hard X-ray and high frequency radio data) found no evidence of such a correlation.

It is clear that we require large samples to shed unambiguous light on this issue. Furthermore, these samples should have uniform datasets in both the radio and X-ray bands, span a large dynamic range, and be obtained simultaneously, so that we can carry meaningful comparisons. However, fulfilling all of these requirements is not feasible at present, mainly due to the lack of sensitive wide field surveys in the two bands.

In the radio bands, the Square Kilometre Array (*SKA*; https://www.skatelescope.org/) is expected to provide a breakthrough in sensitivity starting from its phase 1 in 2020. In its phase 1, the *SKA* will consist of three different components (see Dewdney et al. 2013), one at low frequency (SKA1-LOW, 50–350 MHz), one at intermediate frequency (SKA1-MID, 350 MHz–3 GHz), and one designed for high speed sensitive survey of the sky (SKA1-SUR, 650 MHz–1.7 GHz). Thanks to the wide fraction of accessible sky available with the WFM and the high sensitivity for short exposure available with the LAD, *LOFT* will be an ideal partner for *SKA* in many science areas. In particular, for the physics of accretion in AGN, we can envisage a very powerful synergy between WFM and SKA1-SUR.

Terashima & Wilson (2003) defined a radio loudness parameter $R_X = \nu L_\nu (5\,\text{GHz})/L_X$ and classified radio quiet AGN as those having $\log R_X < -4.5$. Therefore, at the $5\sigma$ WFM sensitivity, about 40 sources in one day and with flux density $\log S_r < \sim -0.5$ Jy at 1.4 GHz would be classified as radio quiet. This sensitivity limit is well within reach of current radio interferometers, even for low values of the radio loudness parameter. Thus, it will be possible to carry out a time-resolved characterization of the source in the WFM field of view on a daily basis. Indeed, while a pointed observation of the single sources could be possible even with present day interferometers, the large (18 deg$^2$ at 1.2 GHz) field of view of SKA1-SUR will permit to carry out a snapshot of the entire WFM daily area. In this way, it will be possible to carry out both a statistical study of the radio–X-ray correlation on a daily basis, and to obtain long series of data for radio–X-ray light curve cross correlations for many sources. The radio emission of radio galaxies and blazars is entirely dominated by synchrotron emission from relativistic jets, further amplified by Doppler beaming in blazars. In particular, at low frequencies ($\nu < \sim 5$ GHz) the radio emission from extended lobes accounts for most of the emitted radiation. On the other hand, the situation in radio quiet sources is much less clear. A nuclear compact component is typically detected but only a few studies have tackled the issue of variability (Barvainis et al. 2005). The joint radio–X-ray (*SKA*-*LOFT*) campaigns described above will be revolutionary in this research area, by providing constraints on the spatial extension and the physical mechanisms at work in both bands (Jones et al. 2011).

### 3.2.3 *LOFT* and the Large Synoptic Survey Telescope (*LSST*)

X-ray and optical variability bears the signature of the physical processes at play in the central parts of the accretion flow in AGN. Combined monitoring campaigns of Seyfert galaxies in the optical/UV and in the X-rays are able to disentangle the physical processes, for instance through the detection of correlations and lags between X-ray and optical/UV light curves. Because of the difficulty in organizing the required monitoring campaigns, only a few objects have been studied up to now and the picture remains rather inconclusive. Although evidence for reprocessing has been found in several objects (e.g., Breedt et al. 2010; Cameron et al. 2012), reprocessing cannot explain all observed features (Arévalo et al. 2008), and no correlation is found in some objects (Maoz et al. 2002). The amplitude of the optical variability compared to that in the X-rays seems also in contradiction with the reprocessing model (Breedt et al. 2009; Arévalo et al. 2008, 2009).

The advent of *LOFT* will change this picture, in fact the WFM will provide well sampled light curves for hundred of AGN (see Fig. 2). In the optical/UV, the Large Synoptic Survey Telescope (*LSST*; http://www.lsst.org) will be in operation at the same time as the planned *LOFT* operations. *LSST* will observe the whole Southern sky from the U photometric band (near ultraviolet) to the Y band (near infrared), with about 100 passes per year on any given object in the visible sky.

This coincidence will benefit enormously the study of physical mechanisms responsible for the blue-bump and for X-ray emission, i.e., the direct signatures of the accretion physics on supermassive BHs. In fact, it will be possible, for the first time, to perform a systematic multiwavelength study on a large sample of Seyfert galaxies, through the connections between optical-UV-X-rays variability in the time domain. The simultaneous optical and X-ray view of AGN will be essential for interpreting the many millions of *LSST* lightcurves of fainter and more distant sources, these coordinated observations will settle the role of X-rays in driving optical and will allow us to disentangle the mass-correlations in optical lightcurves. *LSST* and *LOFT*/WFM joint efforts will open the time domain for massive and accurate studies of varying AGN with unprecedented coverage in flux, wavelength, and timescale.

## 4 Mapping the circumnuclear region around the central supermassive BH

### 4.1 The emitting and absorbing regions: from kpc to sub-pc scale

An Fe K$\alpha$ line and a Compton reflection component are ubiquitous features in the X-ray spectra of Seyfert galaxies (e.g., Perola et al. 2002, Bianchi et al. 2004, 2009). The narrow core of the line is typically unresolved, and must be produced far from the nucleus, either in the BLR, the torus, or the NLR. In order to estimate the geometry and distance of the material responsible for their production, accurate X-ray reverberation analysis of the Fe line and the Compton reflection component represents a promising and powerful technique, by taking into account in detail how the material reacts to the intrinsic variability of the central source. Unfortunately, this is an extremely difficult and uncertain task with current X-ray missions. Continuous light-curves for the continuum can be provided by *Swift*/BAT, but only on very bright sources and on long timescales. Moreover, the Fe line fluxes can be only be sparsely sampled with archival targeted observations from different telescopes, or limited campaigns on selected sources.

*LOFT* will allow us, for the first time, to perform X-ray reverberation of the narrow core of the Fe line with a single mission. The WFM very large field of view is a powerful tool to monitor the continuum intensity of virtually all the sources in the sky. Taking into account the exposure map on one year timescale, the WFM will produce continuous light-curves with $3\sigma$ daily (on average) time-bins for bright sources ($\sim 1 \times 10^{-10}$ erg cm$^{-2}$ s$^{-1}$ in the 7–50 keV band, i.e., above the Fe K edge). Weaker objects ($\sim 5 \times 10^{-11}$ erg cm$^{-2}$ s$^{-1}$) will have $3\sigma$ weekly (on average) time-bins. These timescales are perfectly suited for the Fe narrow line reverberation analysis, since the expected timescales are from days to weeks to years, depending on the location and geometry of the material, and on the intrinsic variability of the sources (therefore on the BH mass and the accretion rate). At the same time, the large area available with the LAD will allow us to perform a well-sampled monitoring for the iron line and Compton reflection component with short targeted observations. The CCD-class energy resolution available with the LAD will easily allow us to disentangle the broad Fe line component from the narrow core. In fact, in only 1 ks, the narrow Fe line flux and reflection fraction can be both recovered with an uncertainty of $\sim 10\%$ for bright sources ($\sim 1 \times 10^{-10}$ erg cm$^{-2}$ s$^{-1}$). The same uncertainties can be recovered for weaker sources ($\sim 5 \times 10^{-11}$ erg cm$^{-2}$ s$^{-1}$) with 5 ks exposures.

Therefore, the synergy between the WFM standard operation and a well-designed monitoring campaign with the LAD will finally allow us to take advantage of X-ray reverberation mapping, to definitely understand the origin of the narrow Fe line and to investigate the physical and geometrical conditions of the emitting regions. Of course, the possibility to adapt for a limited time the WFM observing strategy to well-focused reverberation campaigns would widen the possible outcomes of this technique.

The standard "torus" invoked by Unification Models, in the generic sense of an axisymmetric, rather than spherically symmetric, circumnuclear absorber is certainly present in AGN, but its physical and geometrical structure is quite complex. Indeed, there is now strong evidence of at least three absorption components on very different scales: on scales of hundreds of parsecs, on the parsec scale, and within the dust sublimation radius, on sub-parsec scale (e.g. Bianchi et al. 2012). The most effective way to estimate the distance of the absorber is by means of the analysis of the variability of its column density.

In particular, rapid (from a few hours to a few days) $N_H$ variability has been observed in most bright AGN in the local Universe (Torricelli-Ciamponi et al. 2014), suggesting that obscuration in X-rays is due, at least in part, to BLR clouds. In the "best" case of NGC 1365 such "eclipses" have been observed in detail through wide band X-ray spectroscopy with Suzaku (Maiolino et al. 2010) and jointly with *XMM-Newton* and *NuSTAR* (Risaliti et al. 2013) achieving a precision on the determination of the covering factor of ~5–10% and ~4–8%, respectively. Detailed simulations show that such a measurements (assuming the same observed fluxes) will be obtained with *LOFT* with significantly higher precision (~2%), even in shorter timescales (few tens of seconds with respect to ~100 ks). Moreover, the broad-band energy range offered by *LOFT* will allow us to explore in detail also those clouds with larger column densities, whose variability mostly affects the spectrum at high energies.

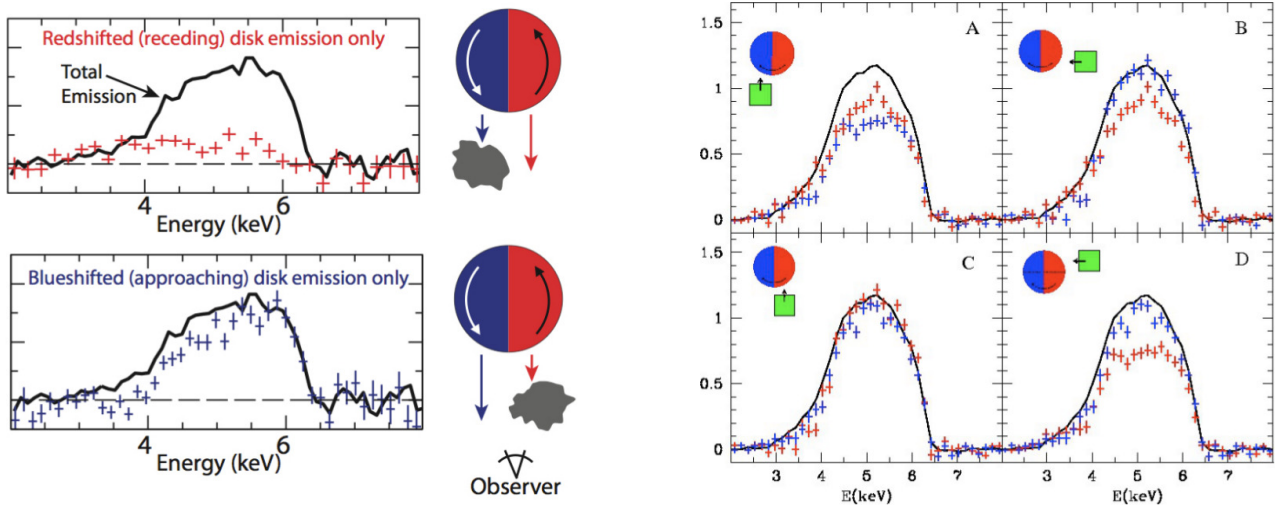

Figure 3: *Left:* simulation of a 50 ks total eclipse by a Compton-thick cloud in NGC 1365. The approaching and receding part of the disc are subsequently eclipsed, allowing the observation of the redshifted and blueshifted line profiles separately. *Right:* The same, for partial eclipses by a Compton-thin cloud, i.e., 50% covering factor and $N_{\rm H} \sim 3 \times 10^{23}$ cm$^{-2}$.

In monitoring observations of 55 Seyferts, *RXTE* detected a dozen eclipse events with durations ranging from $\gtrsim 10$ days to a few months or longer (Markowitz et al. 2014). Such eclipse events are inferred to be from clouds that reside in the outermost BLR or outside the BLR by factors up to ∼10 (the inner regions of infrared-emitting dusty tori). The WFM will be able to follow such events for many more targets with respect to *RXTE* and, crucially, with more regular sampling. Importantly, the combination of simultaneous WFM and *LSST* monitoring will allow us to infer if the eclipsing clouds are dusty or non-dusty via reddening or lack thereof in the optical/UV band (dusty clouds are expected to reside outside the dust sublimation radius, but non-dusty clouds could either be BLR, or be outside the BLR but have some extra source of heating). With WFM observations we can test if dusty torus/ non-dusty BLR clouds are part of the same radial distribution (Elitzur 2007) which, speculatively, may point to their *common* origin, for example in a disc wind.

### 4.2   The nuclear regions: accretion disc and hot corona

A new, potentially decisive way to test the physics of the inner regions of the accretion disc is represented by the detection of variations of the profile of the relativistic iron K$\alpha$ line during the occultations by the BLR clouds described in the previous subsection. In Fig. 3 (left) we show a simulation of a "perfect" eclipse observed by *LOFT*-LAD, due to a Compton-thick cloud completely covering the X-ray source in a few hours. The detection of such an event is possible, though quite rare. This would allow a precise determination of the BH spin, and an unambiguous probe of matter undergoing the effects of General Relativity in the strong-field regime, which is one of the primary scientific objectives of *LOFT*. Perhaps even more interesting is the case shown in the right panel of Fig. 3: these are simulations of the iron line profile observed at different phases of a partial (50% covering factor) eclipse by a Compton-thin cloud ($N_{\rm H} \sim 3 \times 10^{23}$ cm$^{-2}$). These occultations are not rare: they will be observed several times by chance in most bright AGN. The simulations show that *LOFT*, thanks to the wide effective area and CCD-class energy resolution, will be able to measure the variation of the iron profiles even in these quite common events in a large sample of AGN.

The exceptional ability of *LOFT* to obtain an AGN spectrum in a short integration time will allow for direct studies of the spectral energy evolution during strong individual flares (such as the one detected by Ponti et al. 2004), directly in the time domain, by modeling the source's energy spectrum on short timescales. Such a possibility is important since it allows to disentangle the spectral evolution within the emission region from the spectral variability due to the reprocessing. Study of such individual flares in the time domain will be crucial, as they will allow a direct insight into the nature of the "primary emission", i.e., the size, the density, and optical depth, the exact dissipation process (shocks or magnetic field reconnection), and also the location of the event from its own individual reflection component. Studying the extreme values of the energy shift of the relativistic line from the inner regions of a flare-illuminated accretion disc can provide a feasible technique to map the disc surface (Karas et al. 2010; Sochora et al. 2011).

## 5 The BH–galaxy interaction: ultra-fast ionized winds

Today we know that the mass of the supermassive BH mass correlates with several host galaxy properties, indicating co-evolution (e.g., Ferrarese & Merritt 2000). However, how such compact objects can influence galaxy-sized regions still remains one of the mysteries of contemporary astrophysics. BH winds/outflows represent a promising candidate for this link, establishing the so-called "quasar-mode" feedback (e.g., Fabian 2012). Blue-shifted Fe K absorption lines in the X-ray spectra of several AGN suggest the presence of powerful ultra-fast outflows (UFOs) with velocities $\sim 0.1c$ (e.g., Tombesi et al. 2010). These UFOs are observable at locations of sub-parsec scales from the central BH, suggesting an identification with accretion disc winds or the base of a possible weak/broad jet. The associated mass outflow rates of $\sim 1 M_\odot \, \mathrm{yr}^{-1}$ and in particular the high kinetic power (a few per cent of the AGN luminosity) indicate that they could provide an important contribution to AGN feedback (Tombesi et al. 2012). Therefore, in the long term, they could be able to significantly influence the bulge evolution, star formation, BH growth and contribute to the establishment of the observed BH-host galaxy relation, such as the $M$-$\sigma$-correlation (e.g., King 2010).

Despite these significant observational developments, the origin and acceleration mechanism(s) of the ionized absorbers in AGN are still debated. Moreover, large uncertainties still affect the estimates of the mass outflow rates and energetics, the latter being fundamental towards better quantify their influence on AGN feedback and BH-galaxy evolution. Even more fundamental is the capability to observe such transient features with short exposures (few ks) in order to investigate their variability on the dynamical timescale of the absorber. Nevertheless, the statistical significance of the detected absorption lines, ionization states and their variability are still somewhat ambiguous, given the sensitivity limits of the current X-ray satellites for short exposures.

The high effective area and CCD-like energy resolution of *LOFT* LAD can provide a very important contribution to these studies. For instance, in Fig. 4 we show a 20 ks *LOFT* LAD simulation of the Seyfert 1 galaxy Mrk 509 (2–10 keV flux of $\sim 3 \times 10^{-11} \, \mathrm{erg \, cm^{-2} \, s^{-1}}$), which showed UFOs with a velocity of 0.17c in its XMM-Newton spectra (Cappi et al. 2009; Tombesi et al. 2010).

We consider an Fe K$\alpha$ emission line at the rest-frame energy of 6.4 keV and a blue-shifted Fe XXVI absorption line at the observed energies of 8, 10, 12 and 14 keV with an equivalent width of 30 eV. We will be able to clearly detect the UFO up to the higher velocities with a very high significance ($> 5\sigma$) in a timescale of 20 ks. The unique combination of very high effective area peaking at $\sim 10$ keV, moderate energy resolution, and continuous coverage up to energies of $\sim 20$–30 keV will allow a breakthrough in UFO studies compared to current and upcoming satellites, such as *XMM-Newton*, *Chandra*, *Suzaku*, *NuSTAR*, *Astro-H*, and also *Athena*, when considering the AGN in the local Universe with the fastest outflows. In fact, thanks to the high effective area, *LOFT* LAD will be the unique mission able to observe these transient features on short timescales (ks) and, thanks to the large energy band, will allow us to characterize fast outflows (0.6–0.7c) in the local Universe.

In general, *LOFT* LAD will allow this search for UFOs in >40 sources with 2–10 keV fluxes greater that of $\sim 10^{-11} \, \mathrm{erg \, cm^{-2} \, s^{-1}}$, representing the majority of the population of local AGN and, even more fundamentally, to investigate their variability on short timescales. In fact, we will be able to perform short timescale monitoring of the Fe K absorption features and cross-correlate this with the variability in different continuum energy bands. This will allow us to disentangle the main variability driver of the absorber (e.g., ionization, column density, kinematics), as suggested by recent studies of BAL quasars hosting UFOs (e.g., Trevese et al. 2013). The detection of UFOs with velocities $> 0.3c$ could reveal components with the highest mechanical power, and thus potentially having the highest impact on AGN feedback. This will also help us to disentangle the main acceleration mechanism, i.e., radiation or magnetohydrodynamic (MHD) driving. In particular, the detection of velocities up to $0.6$–$0.7c$ could point toward a MHD driving mechanism, possibly similar to that of jets in radio galaxies, given that radiation pressure would have difficulties in producing such values. Finally, the study of UFOs with the highest velocities in radio galaxies (Tombesi et al. 2014) could provide insights into their relation(s) with the relativistic jet. In particular, the detection of UFOs with velocities up to $0.6$–$0.7c$ would indicate that we are mapping the formation/ejection of these outflows very close to the central supermassive BH, where the rotational and escape velocities are highest. They could also represent the outer, mildly-relativistic layers of a stratified jet and could contribute to the jet collimation.

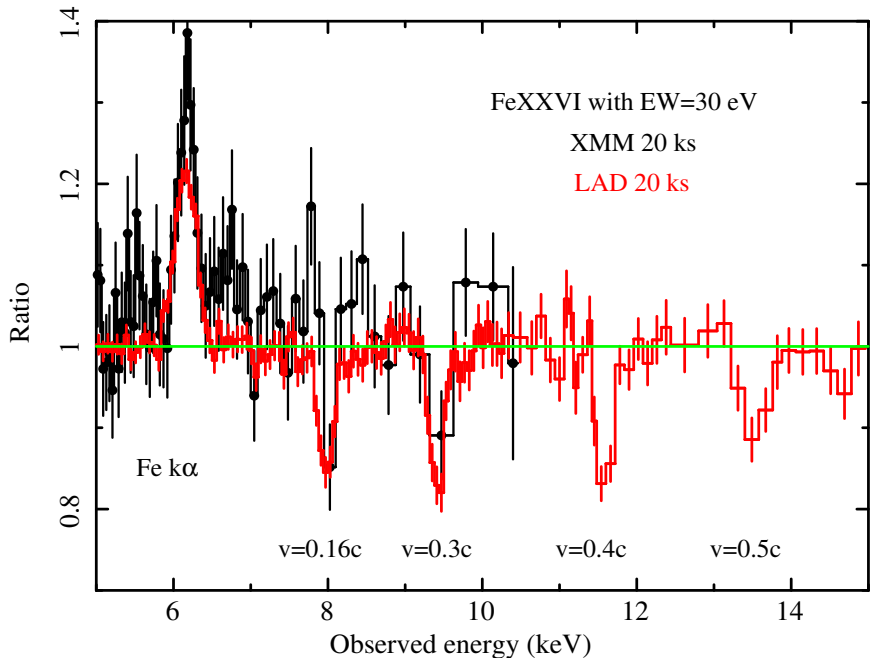

Figure 4: The spectrum of the Seyfert 1 galaxy Mrk 509 as observed with *XMM* and simulated with the *LOFT* LAD (20 ks). An Fe XXVI absorption line between $\sim 8$–$14\,\mathrm{keV}$ is indicative of a highly ionized UFO with an outflow velocity ranging between $0.16 - 0.5c$.